# Geminate recombination dynamics studied via electron reexcitation: Kinetic analysis for anion CTTS photosystems. ¶

Ilya A. Shkrob

[a] *Chemistry Division, Argonne National Laboratory, Argonne, IL 60439*

## Abstract

Recently, it became practicable to study geminate recombination dynamics of solvated electrons in polar liquids by using short pulses of light to reexcite these electrons back into the conduction band of the liquid and observe a change in the fraction of electrons that escape geminate recombination. In this Letter, the potential of this technique to provide additional insight into the recombination dynamics of electrons generated by charge-transfer-to-solvent (CTTS) photodetachment from monovalent anions in polar liquids is studied theoretically. The resulting expression accounts for the recent results for electron photodetachment from $Na^-$ in tetrahydrofuran.

## 1. Introduction

Photoionization of solute/solvent molecules and/or electron photodetachment from anions in molecular liquids results in the formation of short-lived geminate pairs that undergo recombination and diffusional escape to the liquid bulk, typically on the subnanosecond time scale. Studying these dynamics provides useful information on the reactivity of the species involved (e.g., refs. [1-5]). It also yields the initial distribution $P(\mathbf{r})$ of thermalized, localized electrons around their geminate partners, which reflects the photophysics of charge separation on the femtosecond time scale. [1,2,6,7] From the theoretical standpoint, the geminate dynamics are fully described by the probability density $\rho(\mathbf{r};\tau)$ of finding the electron at the point $\mathbf{r}$ in space at the moment $t = \tau$ after the photoexcitation (in the following, the geminate partner of the electron is taken as the origin of the coordinate frame). Typically, these geminate dynamics are observed using pump-probe transient absorption spectroscopy: a short laser pulse is used for photoionization (or electron photodetachment), and second pulse of a different frequency (delayed by $\tau$ relative to the first pulse) is used to obtain the absorption signal from the electrons present at this delay time. [1-7] As a result, one obtains the survival probability $\Omega(\tau)$ of the electrons

$$\Omega(\tau) = \oint_V d^3\mathbf{r}\ \rho(\mathbf{r};\tau) \qquad (1)$$

which is the density $\rho(\mathbf{r};\tau)$ averaged over the reaction volume. While this quantity is important for kinetic analyses, it is much less sensitive to the initial spatial distribution of the electrons than the probability density itself. Thus, a kinetic method which provides additional constraints on this distribution would be very useful.

One of the promising new approaches suggested to obtain additional insight in the recombination dynamics of geminate pairs is to use a second pump pulse to reexcite the solvated electron and promote it into the conduction band of the solvent. The resulting "hot" electron rapidly thermalizes and localizes at some distance away from its original location. On average, this photoexcitation *increases* the separation between the electron and its geminate partner and *suppresses* their recombination, increasing the escape yield of the electron. [7-15] By plotting this yield vs. the delay time $\tau$ of the reexcitation pulse, one obtains a qualitatively different type of kinetics from that given by eq. (1).

For electrons in nonpolar liquids, this technique (known as "photoassisted pair dissociation") [11,12] was first demonstrated by Braun and Scott [7] and further developed by Lukin and co-workers. [11-14] Since the escape yield of the electrons may be determined long after the geminate stage is complete, [7,11] dc conductivity was used to detect these escaped electrons. In these nonpolar liquids, the geminate dynamics of electron-hole pairs are fully controlled by long-range Coulomb interaction between the electron and the hole; [15] the effect of the solvent structure on this interaction is negligible. In polar liquids, the situation is more complex since solvent molecules form a cage around the geminate partners preventing their rapid dissociation. Ultrafast kinetic studies of electrons photolytically detached from their parent anions, such as halides ($Cl^-$, $Br^-$, $I^-$) [1,16,17,18,19] and pseudohalides (e.g., $OH^-$) [2,18,19] in aqueous solutions or alkalide (e.g., $Na^-$) anions in ethers, [4,10,20,21] suggest the formation of short-lived caged pairs after charge-transfer-to-solvent (CTTS) photoexcitation of the corresponding anions. It has been demonstrated experimentally [1,2] that the electron dynamics for aqueous anion photosystems are accounted for by a model in which the electron undergoes diffusion in a central force field corresponding to a mean-force potential (MFP) having a radial profile of a well with the depth of a few *kT* (thermal energy) units.

For polar liquids, transient absorbance of the solvated electron in the visible and near IR may be used to determine the electron concentration at any delay time. Such pump-probe measurements have been routinely performed in the context of a 3-pulse experiment in which two (pump) pulses are used for electron generation and excitation, respectively, and the third (probe) pulse is used for the detection of the electron absorbance. [8,9,10] These experiments have been pursued by several laser spectroscopy groups (e.g., refs. [10,20,22-26]) in order to study the solvation/relaxation dynamics of photoexcited electrons. The modification of this experiment to probe the geminate dynamics of the photoexcited electron is straightforward, and such measurements have been demonstrated by Barbara's and Schwartz's groups: Son et al. [8,9] demonstrated this technique for electrons generated by photoionization of liquid water, whereas Martini and Schwartz [10] used it for electrons generated by Na⁻ CTTS in liquid tetrahydrofuran (THF). The practicability of such 3-pulse experiments poses a question, i.e., what additional inference about the geminate recombination dynamics can be obtained using the reexcitation method?

In this Letter, we give a theoretical analysis of this method for electron dynamics in the anion CTTS photosystems. We seek a model suitable to describe short-range interaction that occurs in such photosystems between the geminate partners. To this end, the result of Lukin et al. [11,12,15] (obtained for Coulomb interaction) is generalized for the general case of an arbitrary MFP; Shushin's theory of the diffusion in the potential well [2,28,29] is then used to obtain the approximate solution. The result is compared with the recent data on electron photodetachment from Na⁻ in THF obtained by Martini and Schwartz. [10]

**2. Results and Discussion.**

In the following, we assume, following Bradforth and coworkers, [1,2] that the dynamics of geminate pairs generated by CTTS excitation of anions can be adequately described by diffusion in a mean force potential $U(\mathbf{r}) = kTu(\mathbf{r})$. In such a situation, the probability density $\rho(\mathbf{r};t)$ obeys the Smoluchowski equation [1,2,27,28,29]

$$\partial \rho / \partial t = D \nabla \bullet [\nabla \rho + \rho \nabla u] \tag{2}$$

or, taking the Laplace transform of both parts of eq. (2),

$$P(\mathbf{r}) - s\tilde{\rho}(\mathbf{r};s) = D \nabla \bullet \left[ e^{-u} \nabla (e^{u} \tilde{\rho}) \right], \tag{3}$$

where $\tilde{\rho}(\mathbf{r};s)$ is the Laplace transform of the density function $\rho(\mathbf{r};t)$ and $P(\mathbf{r})$ is the initial electron distribution. The Green function $\tilde{G}(\mathbf{r},\mathbf{r}';s)$ of eq. (3) is defined by the equation

$$\delta^3(\mathbf{r} - \mathbf{r}') = \left( s + D \nabla \bullet \left[ e^{-u} \nabla (e^{u} ...) \right] \right) \tilde{G}(\mathbf{r},\mathbf{r}';s). \tag{4}$$

The solution of eq. (3) is given by

$$\tilde{\rho}(\mathbf{r};s) = \oint_V d^3\mathbf{r}'\ \tilde{G}(\mathbf{r},\mathbf{r}';s)P(\mathbf{r}'), \tag{5}$$

and, therefore,

$$\rho(\mathbf{r};t) = \oint_V d^3\mathbf{r}'\ G(\mathbf{r},\mathbf{r}';t)P(\mathbf{r}'), \tag{6}$$

where $G(\mathbf{r},\mathbf{r}';t)$ is the inverse Laplace transform of $\tilde{G}(\mathbf{r},\mathbf{r}';s)$. For an electron whose diffusion trajectory starts at the point $\mathbf{r}$, the escape probability

$$\Psi(\mathbf{r}) = \oint_V d^3\mathbf{r}'\ G(\mathbf{r}',\mathbf{r};t=\infty). \tag{7}$$

Observing that the differential equation for the function $\tilde{g}(\mathbf{r},\mathbf{r}';s) = e^{u(\mathbf{r})}\tilde{G}(\mathbf{r},\mathbf{r}';s)$ is self-adjoint, [29] i.e., $\tilde{g}(\mathbf{r},\mathbf{r}';s) = \tilde{g}(\mathbf{r}',\mathbf{r};s)$, eq. (7) may be rewritten as

$$\Psi(\mathbf{r}) = \lim_{s \to 0} \oint_V d^3\mathbf{r}'\ s e^{u(\mathbf{r})-u(\mathbf{r}')}\tilde{G}(\mathbf{r},\mathbf{r}';s). \tag{8}$$

For any delay time $\tau$, the escape probability $\Omega_\infty$ of the electron is given by

$$\Omega_\infty = \oint_V d^3\mathbf{r}\ \Psi(\mathbf{r})\ \rho(\mathbf{r};\tau). \tag{9}$$

Let us assume that at the delay time $t = \tau$ the electron at the point $\mathbf{r}$ is photoexcited and then rapidly trapped by the solvent at the point $\mathbf{r} + \boldsymbol{\xi}$. We will assume that the average displacement $\xi \ll r$. Following this rapid trapping, the escape probability of the reexcited electron at $t = \infty$ is given by

$$\Omega_\infty^*(\tau) = \oint_V d^3\mathbf{r}\ \langle\Psi(\mathbf{r}+\boldsymbol{\xi})\rangle\ \rho(\mathbf{r};\tau), \tag{10}$$

5.

where the asterisk indicates that this probability is for the electron that was photoexcited at $t = \tau$, and $\langle ... \rangle$ stands for averaging over the distribution of displacements $\xi$. Expanding $\Psi(\mathbf{r})$ in the Taylor series, we obtain

$$\Psi(\mathbf{r} + \xi) \approx \Psi(\mathbf{r}) + \sum_l \frac{\partial \Psi(\mathbf{r})}{\partial \xi_l} \xi_l + \frac{1}{2} \sum_{l,m} \frac{\partial^2 \Psi(\mathbf{r})}{\partial \xi_l \partial \xi_m} \xi_l \xi_m + ... \qquad (11)$$

Assuming that the distribution of $\xi$ is isotropic, i.e., $\langle \xi_l \rangle = 0$ and $\langle \xi_l \xi_m \rangle = 1/3 \langle \xi^2 \rangle \delta_{lm}$, the average of eq. (11) is given by

$$\langle \Psi(\mathbf{r} + \xi) \rangle \approx \Psi(\mathbf{r}) + \frac{\Lambda^2}{6} \nabla^2 \Psi(\mathbf{r}), \qquad (12)$$

where $\Lambda^2 = \langle \xi^2 \rangle$ is the mean square displacement of the photoexcited electron. Substituting eq. (12) into eq. (10) and then subtracting eq. (9) from the resulting expression, we obtain the following formula for $\Delta\Omega_\infty(\tau) = \Omega_\infty^*(\tau) - \Omega_\infty$:

$$\Delta\Omega_\infty(\tau) \approx \frac{\Lambda^2}{6} \oint_V d^3\mathbf{r} \ \nabla^2 \Psi(\mathbf{r}) \ \rho(\mathbf{r};\tau) \qquad (13)$$

It is easy to demonstrate, [29] by substituting eq. (7) into eq. (3) and integrating over $\mathbf{r}'$ for $s \to 0$, that the escape probability $\Psi(\mathbf{r})$ obeys the equation

$$\nabla^2 \Psi = \nabla u \ \nabla \Psi \qquad (14)$$

For a central field, this equation has the solution

$$\partial \Psi / \partial r = a/r^2 \exp[u(r)], \qquad (15)$$

in which the parameter $a$ is the Onsager radius of the potential $u(r)$ [27,28] given by

$$a^{-1} = \int dr \ r^{-2} \exp[u(r)], \qquad (16)$$

where the integral is taken over the reaction volume. Substituting eq. (14) into eq. (13) gives

6.

$$\Delta\Omega_\infty(\tau) \approx \frac{\Lambda^2}{6} \oint_V d^3\mathbf{r} \; \nabla u \; \nabla\Psi(\mathbf{r}) \; \rho(\mathbf{r};\tau), \tag{17}$$

or, using the explicit expression for $\partial\Psi/\partial r$ given by eq. (15),

$$\Delta\Omega_\infty(\tau) \approx \frac{a\Lambda^2}{6} \int dr \; 4\pi \; \frac{du}{dr} \; e^{u(r)} \; \rho(r;\tau), \tag{18}$$

which generalizes the expression obtained by Lukin et al. [11,15] for a Coulomb potential. From eq. (18) we obtain that $\Delta\Omega_\infty(\tau) > 0$ for *any* attractive potential. Another important result is that in the absence of interaction between the geminate partners, photoexcitation of the electron has no effect on the escape yield.

In the derivation given above, forced diffusion of the "hot" electron in the conduction band was neglected. [11-15] Let $u_h(\mathbf{r})$ be the potential for this "hot" electron, $D_h$ be its diffusivity and $\tau_h$ be its life time. The mean displacement during the thermalization of the "hot" electron is then given by $\langle\xi\rangle \approx -\tau_h D_h \nabla u_h$. Writing the second term on the right side of eq. (11) as $\xi\nabla\Psi$, we obtain eq. (17) in which $\nabla u$ is replaced by $\nabla\tilde{u} = \nabla u - \beta\nabla u_h$, where $\beta = 6 D_h \tau_h / \Lambda^2$ is a dimensionless parameter.

The expression for $\Delta\Omega_\infty(\tau)$ may be further simplified using Shushin's theory of the diffusion in a potential well. This theory has been used to simulate the electron dynamics following CTTS photoexcitation of several aqueous anions. [2,30] In this theory, the solution $\rho(r;t)$ of eq. (2) is obtained by splicing two density functions: (i) the population of the well $n(t)$ (which includes all geminate pairs for which $r \leq a$) and (ii) the exact solution of eq. (2) for $u(r) = 0$ at $r \geq a$ (see refs. [2], [27] and [28] for more discussion of this model). Since for $r > a$ $\nabla u \approx 0$, the integral in eq. (17) may be estimated for $r \approx a$ only. In Shushin's theory,

$$\Psi(r) = 1 - (1 - p_d) \; a/r \tag{19}$$

for $r > a$, [2] where $p_d = W_d/W$ is the escape probability for a geminate pair generated inside the potential well (i.e., for $r \leq a$) and $W = W_d + W_r$ is the sum of the rate constants for escape and recombination of caged pairs, respectively. [27,28] From eq. (19), we obtain $(d\Psi/dr)_{r=a} = (1-p_d)/a$. Substituting this identity into eq. (17), we obtain

$$\Delta\Omega_\infty(\tau) \approx \frac{\Lambda^2}{6a^2}\left(r\frac{du}{dr}\right)_{r=a}\frac{W_r}{W} n(\tau). \tag{20}$$

In Shushin's model, [28] the survival probability $\Omega(t)$ of the geminate pair (eq. (1)) is given by

$$d\Omega(t)/dt = -W_r n(t). \tag{21}$$

Substituting the latter equation into eq. (20), we finally obtain (for $\Lambda << a$),

$$\Delta\Omega_\infty(\tau) \approx \frac{\Lambda^2}{6Wa^2}\left(r\frac{du}{dr}\right)_{r=a}\left[-\frac{d\Omega(\tau)}{d\tau}\right]. \tag{22}$$

Thus, the change $\Delta\Omega_\infty(\tau)$ in the escape probability induced by reexcitation of the electron at $t = \tau$ is proportional to the derivative of the survival probability $\Omega(\tau)$ at this delay time. We conclude that for a compact mean force potential, the 3-pulse experiment yields much the same information as the 2-pulse experiment (see the Introduction). For the electron dynamics in a Coulomb field [11-15] (or any other diffuse field), no such obvious correlation between these two types of kinetics exists.

In Fig. 1, eq. (22) is put to test using the recent experimental data of Martini and Schwartz for electron photodetachment from Na⁻ in liquid THF. [10] The sodide anion was photoexcited by a short pulse of 0.79 μm light and the resulting trapped photoelectron excited into the conduction band by a 2 μm laser pulse. The change $\Delta\Omega_\infty(\tau)$ in the escape yield of the electrons due to the 2 μm light photoexcitation was determined from the (solvated) electron absorbance at 1.25 μm for $t$=500 ps and plotted as a function of the time interval $\tau$ between the 0.79 μm and 2 μm laser pulses (trace (i)). In a separate experiment, transient absorbance of the electron at 2 μm induced by the 0.79 μm pulse was determined (trace (ii)) yielding the survival probability $\Omega(\tau)$. In Fig. 1, the derivative of the latter curve and the function $\Delta\Omega_\infty(\tau)$ are compared. These two traces appear to be very similar, supporting our general conclusion.

While this good correspondence between the experiment and the theory is reassuring, it would be desirable to test eq. (22) for an *aqueous* anion photosystem. In the sodide/THF photosystem, the relaxation of photoexcited electron (after its subsequent localization by the solvent) occurs on the same time scale as the separation of caged $(Na^{\bullet}, e_s^-)$ pair (although the experiments of Martini and Schwartz [10,31] suggest that the kinetics of this relaxation does not change as a function of the delay time of the 2 μm pulse). In water, the dissociation of caged pairs in the anion CTTS photosystems is an order of magnitude longer [1,2,16-19] and the break between the electron solvation (occurring in < 1 ps) [32,33] and the dissociation of caged pairs is more clear. Such an experiment would be worthwhile even if no additional insight into the electron dynamics for the CTTS photosystems is obtained: The demonstration of the correctness of eq. (22) would lend additional support to the potential well model suggested for these dynamics by Bradforth and coworkers. [1,2,30]

**6. Acknowledgement.**

IAS thanks Dr. I. B. Martini and Prof. B. J. Schwartz of UCLA for the permission to reproduce their unpublished data. The research at the ANL was supported by the Office of Science, Division of Chemical Sciences, US-DOE under contract number W-31-109-ENG-38.

**Figure caption.**

**Figure 1.**

Normalized survival probability $\Omega(\tau)$ *(open circles, to the right)* and the increase $\Delta\Omega_\infty(\tau)$ in the fraction of escaped electrons *(filled circles, to the left)* for the $(Na^\bullet, e_s^-)$ pair generated by 0.79 µm photoexcitation of Na⁻ in room-temperature liquid THF. The time profile of the excitation pulse is shown by the dashed line (trace (iii)). See the text and the caption to Fig. 3A in ref. [10] for more detail. The electron is reexcited by a femtosecond pulse of 2 µm light at the delay time $\tau$. To obtain the survival probability, transient absorbance of the solvated electron was determined using this 2 µm pulse as a probe. To obtain $\Delta\Omega_\infty(\tau)$, the electron absorbance (1.25 µm) at *t*=500 ps was plotted as a function of the delay time between the 0.79 µm and 2 µm pulses. The solid line drawn through the open circles is a least squares biexponential fit (trace (ii)). The normalized derivative of this curve (trace (ii)) is juxtaposed onto the data for $\Delta\Omega_\infty(\tau)$.

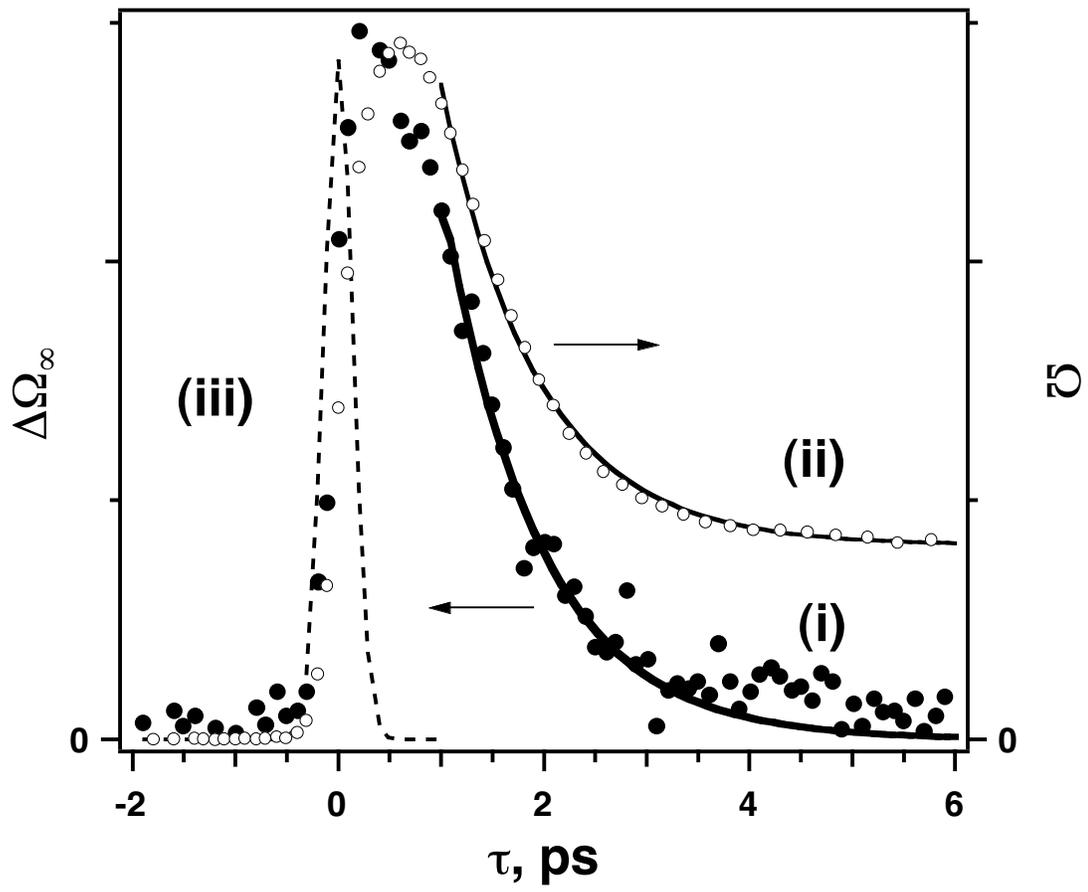

**Figure 1. Shkrob**